\documentclass[aps,prd,10pt,
               preprintnumbers,numbers,sort&compress,nofootinbib,
               twocolumn,
               showpacs,
               colorlinks,
               linkcolor=blue,
               citecolor=blue]{revtex4}

\usepackage{amsmath}
\usepackage{amssymb}
\usepackage{mciteplus}
\mciteErrorOnUnknownfalse

\newcommand{\vect}[1]{\vec{\mathbf{#1}}}
\newcommand{\abs}[1]{|#1|}
\newcommand{\braket}[1]{\langle{#1}\rangle}
\newcommand{\diff}[3][{}]{\frac{\mathrm{d}^{#1}{#2}}{\mathrm{d}{#3}^{#1}}}
\providecommand{\d}{}
\renewcommand{\d}{{\rm d}}

\newcommand{\Chandra}{\textit{Chandra}}
\newcommand{\SPI}{\textsc{spi}}
\newcommand{\CGRO}{\textsc{crgo}}
\newcommand{\INTEGRAL}{\textsc{integral}}
\newcommand{\COMPTEL}{\textsc{comptel}}
\newcommand{\EGRET}{\textsc{egret}}
\newcommand{\WMAP}{\textsc{wmap}}
\newcommand{\QCD}{\textsc{qcd}}
\newcommand{\QED}{\textsc{qed}}
\newcommand{\DM}{\textsc{dm}}
\newcommand{\VM}{\textsc{vm}}
\newcommand{\keV}{\text{ keV}}
\newcommand{\GeV}{\text{ GeV}}
\newcommand{\MeV}{\text{ MeV}}
\newcommand{\eV}{\text{ eV}}

\newcommand{\GHz}{\text{ GHz}}

\newcounter{myenumi}
\newenvironment{myenumerate}[1][]{
  \begin{list}{(#1\arabic{myenumi})}{
      \usecounter{myenumi}
      \setlength\parsep{0pt}
      \setlength\itemsep{0pt}
      \setlength\itemindent{0pt}
      \setlength\topsep{0pt}
      \setlength\listparindent\parindent
      \setlength\labelsep\leftmargin
      \addtolength\labelsep{-\labelwidth}
      \addtolength\labelsep{\itemindent}
    }
}{\end{list}}

\begin{document}
\title{WMAP Haze: Directly Observing Dark Matter?}
\author{Michael McNeil Forbes}
\affiliation{Nuclear Theory Group, Department of Physics, University
  of Washington, Seattle, WA 98195-1560, USA} 
\author{Ariel R. Zhitnitsky}
\affiliation{The University of British Columbia, Department of Physics
  and Astronomy, Vancouver, B.C., V6T 1Z1, Canada}
\begin{abstract}
  In this paper we show that dark matter in the form of dense
  matter/antimatter nuggets could provide a natural and unified
  explanation for several distinct bands of diffuse radiation from the
  core of the Galaxy spanning over 12 orders of magnitude in
  frequency.  We fix all of the phenomenological properties of this
  model by matching to x-ray observations in the keV band, and then
  calculate the unambiguously predicted thermal emission in the
  microwave band, at frequencies smaller by 10 orders of magnitude.
  Remarkably, the intensity and spectrum of the emitted thermal
  radiation are consistent with---and could entirely explain---the
  so-called ``\WMAP\ haze'': a diffuse microwave excess observed from
  the core of our Galaxy by the Wilkinson Microwave Anisotropy Probe
  (\WMAP).  This provides another strong constraint of our proposal,
  and a remarkable nontrivial validation.  If correct, our proposal
  identifies the nature of the dark matter, explains baryogenesis, and
  provides a means to directly probe the matter distribution in our
  Galaxy by analyzing several different types of diffuse emissions.
\end{abstract}
\preprint{NT@UW-08-05}
\maketitle
\tableofcontents
\section{Introduction}
\label{sec:introduction}
In this paper we discuss a testable and well-constrained model for
dark matter~\cite{Zhitnitsky:2002qa,Oaknin:2003uv,Zhitnitsky:2006vt,Forbes:2006ba}.
In particular, we explain how microwave emissions are an inevitable
consequence of our proposal, and test the model against a recent
analysis of the Wilkinson Microwave Anisotropy Probe (\WMAP)
observations that suggest an anomalous emission from the core of our
Galaxy (dubbed the ``\WMAP\
haze'')~\cite{Finkbeiner:2003im,Finkbeiner:2004je,Finkbeiner:2004us,Hooper:2007gi,Dobler:2007wv,Hooper:2008zg}.
Despite having no free parameters---the model is completely fixed by
observations at scales some 10 orders higher---our proposal is
consistent with these observations, and could even explain the anomaly
if it survives further scrutiny.  This provides a highly nontrivial
test of our proposal, which remains consistent with all known
constraints.

To calculate the emissions, however, requires a careful analysis of
several fields of physics, and it is easy to lose track of the overall
structure of the calculation.  We organize the paper as follows: In
section~\ref{sec:proposal}, we present a short review of the
dark-matter proposal, emphasizing the assumptions that underlie the
model, the observational constraints that the model must satisfy, and
outlining the mechanism by which the dark matter will radiate, thus
rendering it observable.  In section~\ref{sec:therm-prop-nugg}, we
summarize the calculation of the thermal microwave emission from the
dark matter, and show in section~\ref{sec:comp-wmap-haze} how this is
consistent with the observations.  Finally, in
section~\ref{sec:summary}, we review all of the observational
constraints of our proposal and discuss the testable predictions that
it makes.  To keep the logic clear, some technical details have been
omitted from core of the paper.  We include these in the appendix,
completing our calculation.
clearpage

\section{Proposal}
\label{sec:proposal}
\subsection{Dark Matter as Dense Quark Nuggets}
\label{sec:dark-matter-as}
Two of the outstanding cosmological mysteries---the natures of dark
matter and baryogenesis---might be explained by the idea that dark
matter consists of Compact Composite Objects
(\textsc{cco}s)~\cite{Zhitnitsky:2002qa,Oaknin:2003uv,Zhitnitsky:2006vt,Forbes:2006ba}
similar to Witten's strangelets~\cite{Witten:1984rs}.  The basic idea
is that these \textsc{cco}s---nuggets of dense matter and
antimatter---form at the same \textsc{qcd} phase transition as
conventional baryons (neutrons and protons), providing a natural
explanation for the similar scales $\Omega_{\DM} \approx
5\Omega_{\textsc{b}}$.  Baryogenesis proceeds through a charge
separation mechanism: both matter and antimatter nuggets form, but the
natural \textsc{cp} violation of the so-called $\theta$ term in
\textsc{qcd}\footnote{If $\theta$ is nonzero, one must confront the
  so-called strong \textsc{cp} problem whereby some mechanism must be
  found to make the effective $\theta$ parameter extremely small today
  in accordance with measurements.  This problem remains one of the
  most outstanding puzzles of the Standard Model, and one of the most
  natural resolutions is to introduce an axion field.  (See the
  original papers
  \cite{Peccei:1977ur,*Weinberg:1978ma,*Wilczek:1978pj},
  \cite{Kim:1979if,*Shifman:1980if},
  \cite{Dine:1981rt,*Zhitnitsky:1980tq}, and recent reviews
  \cite{Srednicki:2002ww,*vanBibber:2006rb,*Asztalos:2006kz}.)  Axion
  domain walls associated with this field (or ultimately, whatever
  mechanism resolves the strong \textsc{cp} problem) play an important
  role in forming these nuggets, and may play in important role in
  their ultimate stability.
  See~\cite{Zhitnitsky:2002qa,Oaknin:2003uv} for details.}---which was
of order unity $\theta\sim 1$ during the \QCD\ phase
transition---drives the formation of more antimatter nuggets than
matter nuggets, resulting in the leftover baryonic matter that forms
visible matter today (see \cite{Oaknin:2003uv} for details).  Note, it
is crucial for our mechanism that \textsc{cp} violation can drive
charge separation.  This idea may already have found experimental
support through the Relativistic Heavy Ion Collider (\textsc{rhic}) at
Brookhaven \cite{Kharzeev:2007tn}.

This mechanism requires no fundamental baryon asymmetry to explain the
observed matter/antimatter asymmetry.  From this, and the observed
relation $\Omega_{\DM} \approx 5\Omega_{\textsc{b}}$
(see~\cite{Amsler:2008zz} for a review) we have
\begin{subequations}
  \label{eq:321}
 \begin{align}
    B_{\text{universe}} = 0 &= B_{\text{nugget}}
    + B_{\text{visible}}-\bar{B}_{\text{antinugget}}\\
    B_{\text{dark-matter}} &= B_{\text{nugget}} +
    \bar{B}_{\text{antinugget}} \approx 5B_{\text{visible}}
  \end{align}
\end{subequations}
where $B_{\text{universe}}$ is the total number of baryons
\emph{minus} the number of antibaryons in the universe,
$B_{\text{dark-matter}}$ is the total number of baryons \emph{plus}
the total number of antibaryons hidden in the nuggets and
antinuggets that make up the dark matter, $B_{\text{nugget}}$ is the
total number of baryons contained in all of the dark-matter nuggets,
$\bar{B}_{\text{antinugget}}$ is the total number of antibaryons
contained in all of the dark-antimatter nuggets, and
$B_{\text{visible}}$ is the total number of residual ``visible''
baryons (regular matter).  Solving equations~(\ref{eq:321}) gives the
approximate ratios
$\bar{B}_{\text{antinugget}}$:$B_{\text{nugget}}$:$B_{\text{visible}}\simeq
$ 3:2:1.

Unlike conventional dark matter candidates, dark-matter/antimatter
nuggets will be strongly interacting, but macroscopically large
objects.  They do not contradict any of the many known observational
constraints on dark matter or
antimatter~\cite{Zhitnitsky:2006vt,Forbes/Zhitnitsky:2007b} for three
reasons:
\begin{myenumerate}
\item They carry a huge (anti)baryon charge $|B| \approx 10^{20}$ --
  $10^{33}$, so they have an extremely tiny number density.  This
  explains why they have not been directly observed on earth.  The
  local number density of dark-matter particles with these masses is
  small enough that interactions with detectors are exceedingly rare
  and fall within all known detector and seismic
  constraints~\cite{Zhitnitsky:2006vt}.  (See also
  \cite{Herrin:2005kb,Abers:2007ji} and references
  therein.\footnote{It is estimated in \cite{Herrin:2005kb} that
    nuggets of mass from $\sim$ 10 kg to 1 ton (corresponding to $B
    \sim 10^{28\text{-}30}$) must account for less than an order of
    magnitude of the local dark matter.  While our preferable range of
    $B\sim 10^{25\text{-}27}$ is somewhat smaller
    \cite{Forbes/Zhitnitsky:2007b} and does not contradict
    \cite{Herrin:2005kb}, we still believe that $B\geq 10^{28}$ is not
    completely excluded by Apollo data, as the corresponding
    constraint is based on specific model dependent assumptions about
    the nugget mass-distribution \cite{Herrin:2005kb}, whereas nugget
    formation due to charge separation as suggested in
    \cite{Oaknin:2003uv} may lead to a very different
    distribution.})
\item The nuggets have nuclear densities, so their interaction cross-section
  is small $\sigma/M \approx 10^{-13}$-$10^{-9}$~cm$^2$/g.  This is
  well below typical astrophysical and cosmological limits which are
  on the order of $\sigma/M<1$~cm$^2$/g.  Dark-matter--dark-matter
  interactions between these nuggets are thus negligible.
\item They have a large binding energy such that baryons in the
  nuggets are not available to participate in big bang nucleosynthesis
  (\textsc{bbn}) at $T \approx 1\MeV$.  In particular, we suspect that
  the baryons in these nuggets form a superfluid with a gap of the
  order $\Delta \approx 100\MeV$, and critical temperature $T_{c} \sim
  \Delta/\sqrt{2} \approx 60$ MeV, as this scale provides a natural
  explanation for the observed photon to baryon ratio
  $n_{\textsc{b}}/n_{\gamma} \sim 10^{-10}$~\cite{Oaknin:2003uv}, which
  requires a formation temperature of $T_{\text{form}} =
  41$\MeV~\cite{kolb94:_early_univer}.\footnote{At temperatures below
    the gap, incident baryons with energies below the gap would
    Andreev reflect rather than become incorporated into the nugget.}
\end{myenumerate}
Thus, on large scales, the nuggets are sufficiently dilute that they
behave as standard collisionless cold dark matter (\textsc{ccdm}).
When the number densities of both dark and visible matter become
sufficiently high, however, dark-antimatter--visible-matter collisions
may release significant radiation and energy.  In particular,
antimatter nuggets provide a site at which interstellar baryonic
matter---mostly protons and electrons---can annihilate, producing
emissions with calculable spectra and energies that should be
observable from the core of our Galaxy.  These emissions are not only
consistent with current observations, but seem to naturally explain
several mysterious diffuse emissions observed from the core of our
Galaxy, with frequencies ranging over some 12 orders of magnitude.

Although somewhat unconventional, this idea naturally explains several
coincidences, is consistent with all known cosmological constraints,
and makes definite, testable predictions.  Furthermore, this idea is
almost entirely rooted in conventional and well-established physics.
In particular, there are no ``free parameters'' that can be---or need
to be---``tuned'' to explain observations: In principle, everything
is calculable from well-established properties of \QCD\ and \QED.  In
practice, fully calculating the properties of these nuggets requires
solving the fermion many-body problem at strong coupling, so we must
resort to ``fitting'' a handful of phenomenological parameters from
observations.

Nevertheless, these unknown parameters may be determined to within an
order of magnitude by observations of processes at the keV scale
(described below and in \cite{Forbes:2006ba}).  The model then makes
unambiguous predictions about other processes ranging over more than
10 orders of magnitude in scale.  The point of this paper is to show
that, remarkably, these unambiguous predictions are completely
consistent with current observations, providing compelling evidence
for our proposal, and explaining another astrophysical puzzle: the
origin of the so-called ``\WMAP\
haze''~\cite{Finkbeiner:2003im,Finkbeiner:2004je,Finkbeiner:2004us,Hooper:2007gi,Dobler:2007wv,
  Hooper:2008zg}.\footnote{We should remark here that our explanation
  of the \WMAP\ haze with dark matter is not a new idea.  It was
  suggested previously that self-annihilating weakly interacting
  massive particles (WIMPs) might explain the \WMAP\ haze \cite{
    Finkbeiner:2004us, Hooper:2007gi, Hooper:2008zg}.  These WIMPs
  must be very heavy, $m\sim 100$ GeV, and therefore, annihilation
  must produce significant amounts of high energy
  radiation~\cite{Bottino:2008sv} if it is to also explain microwave
  emissions with a typical frequency of $\omega\sim 10^{-4}$.  For
  example, if one takes central values for the WIMP parameters, then
  the microwave intensity from WIMP annihilation would be well below
  the observed intensity~\cite {Bottino:2008sv}. In any case, this
  proposal has very different predictions than ours.}  We have
considered five independent observations of diffuse radiation from the
core of our Galaxy:
\begin{myenumerate}
\item S\textsc{pi}/\INTEGRAL\ observes 511\keV\ photons from
  positronium decay that is difficult to explain with conventional
  astrophysical positron
  sources~\cite{Knodlseder:2003sv,Beacom:2005qv,Yuksel:2006fj}.
  Dark-antimatter nuggets would provide an unlimited source of
  positrons as suggested in~\cite{Oaknin:2004mn,Zhitnitsky:2006tu}.
\item C\textsc{omptel}\ detects a puzzling excess of 1--20 MeV $\gamma$-ray
  radiation.  We shall not discuss this here, but it has been shown in
  \cite{Lawson:2007kp} that the direct e$^{+}$e$^{-}$ annihilation
  spectrum could nicely explain this deficit.
\item \Chandra\ observes a diffuse keV \textsc{x}-ray emission that greatly
  exceeds the energy from identified sources~\cite{Muno:2004bs}.
  Visible-matter/dark-antimatter annihilation would provide this
  energy.
\item E\textsc{gret}/\CGRO\ detects MeV to GeV gamma-rays, constraining
  antimatter annihilation rates.  We shall not discuss these
  constraints here, but it was shown in~\cite{Forbes:2006ba} that
  these constraints are consistent with the rates inferred from the
  other emissions.
\item W\textsc{map} has detected an excess of GHz microwave
  radiation---dubbed the ``\WMAP\ haze''---from the inner $20^\circ$
  core of our
  Galaxy~\cite{Finkbeiner:2003im,Finkbeiner:2004je,Finkbeiner:2004us,Hooper:2007gi,Dobler:2007wv}.
  Annihilation energy not immediately released by the above mechanisms
  will thermalize, and subsequently be released as thermal
  bremsstrahlung emission at the eV scale.  Although the eV scale
  emission will be obscured by other astrophysical sources, the tail
  of the emission spectrum is very hard, and carries enough energy in
  the microwave to explain the \WMAP\ haze.
\end{myenumerate}
To proceed, we start with the three basic postulates assumed
in~\cite{Forbes:2006ba}:
\begin{myenumerate}
\item[(A.1)] The antimatter nuggets provide a virtually unlimited source of
  positrons (e$^{+}$) such that impinging electrons (e$^{-}$) will
  readily annihilate at their surface through the formation of
  positronium~\cite{Oaknin:2004mn,Zhitnitsky:2006tu}.
\end{myenumerate}
About a quarter of the positronium annihilations release back-to-back
511 keV photons.  On average one of these photons will be absorbed
by the nugget while the other will be released.
\begin{myenumerate}
\item[(A.2)] The nuggets provide a significant source of
  antibaryonic matter such that impinging protons will annihilate.
  We assume that the proton annihilation rate is directly related to
  that of electrons through a suppression factor $f<1$ as discussed in
  \cite{Forbes:2006ba}.
\end{myenumerate}
Proton annihilation events will release about $2\,m_{p} \approx 2$\GeV\
of energy per event and will occur close to the surface of the nugget
creating a hot spot that will radiate x-ray photons with keV energies
containing some fraction $g$ of the total annihilation energy.  The
remaining fraction $1-g$ will be released at the eV scale, after the
energy has thermalized within the nuggets.  The tail of this thermal
emission will be released in the microwave spectrum and may explain
the observed \WMAP\ haze.  To further test this theory and connect all
of these emissions, we make an additional assumption:
\begin{myenumerate}
\item[(A.3)]  We assume that the emitted 511 keV photons dominate the
  observed 511 keV flux, that the emitted keV x-rays dominate the
  observed diffuse x-ray flux, and that the thermally emitted
  microwaves dominate the observed \WMAP\ haze.
\end{myenumerate}
The basis for this assumption is that none of these fluxes has a
convincing explanation.  The nuggets may thus provide the missing
explanation in each case.  This assumption allows us to use the
observations at the keV scale---the 511\keV\ emission measured by
\INTEGRAL, and the diffuse x-ray emission measured by \Chandra---to
fix all of the phenomenological parameters \cite{Forbes:2006ba}.  The
model then makes unambiguous predictions about the properties of the
\WMAP\ haze, allowing it to be tested.

As we shall see, the agreement is remarkable: even though our
estimates are only accurate up to the order of magnitude, the picture
that dark matter consists mostly of antimatter nuggets can completely
explain all three of these puzzling emissions without contradicting a
single observation.
\subsection{Observable Dark Matter: Emissions}
\label{sec:prop-expl-emiss}
As discussed in~\cite{Forbes:2006ba}, our proposal is that both
electrons and protons annihilate on antimatter nuggets, releasing
observable radiation from ``hot spots'' near the annihilation sites.
These emissions are best though of as ``jets'', and occur sufficiently
rapidly that they are produced on a per-event basis, thus producing a
spectrum that is independent of the local
environment~\cite{Forbes:2006ba}.  This includes the $511\keV$
spectrum from positronium
annihilation~\cite{Oaknin:2004mn,Zhitnitsky:2006tu}, the spectrum up
to $20\MeV$ from direct e$^{+}$e$^{-}$
annihilation~\cite{Lawson:2007kp}, the diffuse $\sim 10\keV$ radiation
from p$^{+}$p$^{-}$ annihilation~\cite{Forbes:2006ba}, and the
occasional GeV photon produced directly from proton
annihilation~\cite{Forbes:2006ba}.

The rates of annihilation and the energies released have been
correlated~\cite{Forbes:2006ba} with the observed diffuse $511\keV$
positronium
emission~\cite{1978ApJ...225L..11L,Kinzer:2001ba,Knodlseder:2003sv,Jean:2005af,Knodlseder:2005yq,Weidenspointner:2006nu}
and diffuse $\sim 10\keV$ emissions~\cite{Muno:2004bs} observed from
the core of the Galaxy, providing a test of the model.  It was shown
that both of these emissions could be nicely accounted for if the rate
of x-ray energy released from p$^{+}$p$^{-}$ annihilation was related
to that of e$^{-}$e$^{+}$ annihilation through a suppression factor
$f\cdot g \sim 6 \times 10^{-3}$ where the factor $f$ accounts for
proton reflection from the sharp nuclear matter interface, and the
factor $\tfrac{1}{10} < g < \tfrac{1}{2}$ accounts for the fraction of
the $2\GeV$ p$^{+}$p$^{-}$ annihilation energy released at the
``hot-spots''.

The topic of this paper is the remaining fraction $1-g$ of the $2\GeV$
annihilation energy that will be transmitted deep withing the nuggets,
ultimately being thermally radiated at a much lower energy
scale.  We shall show that most of this energy will be released at the
eV scale, making it difficult to observe against the bright stellar
background.  The spectrum of this emission, however, will be shown to
be extremely hard, resulting in a significant release of detectable
microwave energy ($\sim 10^{-4}$ eV).

Our main point is that this microwave
emission could fully account for the recently observed
\WMAP\ haze~\cite{Finkbeiner:2003im,Finkbeiner:2004je,Finkbeiner:2004us,Hooper:2007kb}:
a puzzling diffuse emission from the core of our Galaxy.

In Section~\ref{sec:therm-emiss-from} we estimate the thermal
emissivity of nuggets, and the spectrum of the emitted radiation.  In
Section~\ref{sec:therm-nugg} we show how the nuggets reach
thermodynamic equilibrium at an eV scale by balancing the annihilation
rate with the emission.  Armed with these estimates, in
Section~\ref{sec:comp-wmap-haze} we compare the predictions of our
model with the observations of the \WMAP\ haze, using our previous
results~\cite{Forbes:2006ba} to provide the normalizations, and arrive
at the remarkable conclusion that our proposal naturally explains the
energy budget~\ref{sec:energy-budget} and spectrum~\ref{sec:spectrum}
of the observations, even though the predictions are at an energy
scale some 10 orders of magnitude smaller than the 511 keV scale at which the
normalization was fixed!  Finally, in Section~\ref{sec:conclusion} we
reiterate the testable predictions our model makes, thus providing a
method with which to confirm or rule out the proposal over the next
few years.

\section{Thermal Properties of the Nuggets}
\label{sec:therm-prop-nugg}
\subsection{Emissivity}
\label{sec:therm-emiss-from}
Here we discuss the properties of thermal emission from the
``electrosphere'' of the nuggets at low temperatures $T\sim$ eV.  As
we shall show in Section~\ref{sec:therm-nugg}, this temperature can be
established by comparing the rate of annihilation energy deposited in
the nugget with the rate of emission.  In what follows, we shall
present a simple estimate to capture the order of magnitude of the
process.  In principle, the exact numerical factors can be computed,
but such a calculation is extremely tedious, and would be of no use
since there are other uncertainties in this problem of a similar
magnitude.

The emissivity depends on the density $n(z)$ of the positron cloud,
which varies as a function of the distance $z$ from the quark matter
core.  At the eV scale temperatures, the most important region of
emission will be the region of the electrosphere where the kinetic
energy of the particles $p^2/2m \simeq T$ is on the same order as the
temperature.  Closer to the core of the nugget, a well-defined Fermi
surface develops with $p_{F}\gg T$ , and the low-energy excitations
that can scatter and radiate are confined to an effectively
two-dimensional region of momentum space about the Fermi surface.  As
a result, there is a kinematic suppression of the emissivity from
these regions and the emission will not change the order of magnitude
estimate we present here.  Sufficiently deep into the nugget, the
plasma frequency will also be large enough that the emitted eV scale
photons will be highly virtual and thus rapidly reabsorbed.  A
detailed discussion of the suppression of emission from the highly
dense regions is presented in appendix~\ref{sec:dense-regions}.

Here we shall estimate the emissivity of a Boltzmann gas of positrons.
The Boltzmann approximation is valid where $n\ll p^{-3}\sim
(mT)^{3/2}$ and we can neglect both the fermion degeneracy that will
suppress the emissivity closer to the core and many-body effects.
Thus we start from the following expression\footnote{Expression
  (\ref{eq:sigma}) should be contrasted with the well known dipole
  type of expression for \emph{different} types of particles emitting
  soft photons, such as electrons and ions.  With identical particles
  having the same charge to mass ratio $e/m$, the dipole contribution
  is zero, and the cross section is dominated by the quadrupole
  interaction.  This quadrupole character explains the appearance of
  the velocity $\braket{v}$ in numerator of (\ref{eq:spectrum}) as
  opposed to the factor $\braket{v}^{-1}$ that enters the
  corresponding expression for electron-ion collisions.} for the cross
section for two positrons emitting a photon with $\omega\ll p^2/(2m)$
\cite{Fedyushin:1974mt,*achiezer69:_quant_elect,
  *1998SoPh..178..341H},
\begin{equation}
  \label{eq:sigma}
  \d\sigma_{\omega}=\frac{4}{15}\alpha \left(\frac{\alpha}{m}\right)^2
  \cdot\left(17+12\ln\frac{p^2}{m\omega}\right)\frac{\d\omega}{\omega}.
\end{equation}
The emissivity $Q=\d{E}/\d{t}/\d{V}$---defined as the total energy
emitted per unit volume, per unit time---and the spectral properties
can be calculated from
\begin{multline}
  \label{eq:spectrum}
  \frac{\d{Q}}{\d{\omega}}(\omega,z)
  = n_1(z, T)n_2(z, T) 
  \omega
  \left\langle v_{12}\frac{d\sigma_{\omega}}{d\omega}\right\rangle\\
  =\frac{4\alpha}{15}\left(\frac{\alpha}{m}\right)^2n^2(z,T)
 \left\langle v_{12}\left(17+12\ln\frac{p_{12}^2}{m\omega}\right)\right\rangle
\end{multline}
where $n(z,T)$ is the local density at distance $z$ from the nugget's
surface, and $v_{12} = \abs{\vec{v}_1-\vec{v}_2}$ is the relative
velocity.  The velocity and momentum $p_{12}$ need to be thermally
averaged.  To estimate this, we use the Boltzmann ensemble at
temperature $T$ with a kinematic cutoff $p^2/(2m) > \omega$, as only
particles with sufficient energy can emit photons with energy
$\omega$.  In principle one can do better, but the current approach
suffices to give the correct order of magnitude (see
Appendix~\ref{sec:boltzmann-average} for details):
\begin{multline}
  \left\langle v_{12}\left(17+12\ln\frac{mv_{12}^2}{\omega}\right)\right\rangle
  \approx \\
  \approx
  2\sqrt{\frac{2T}{m\pi}}
  \left(1+\frac{\omega}{T}\right)e^{-\omega/T}h\left(\frac{\omega}{T}\right)
\end{multline}
where (this approximation is accurate to about $25\%$)
\begin{equation}
  h(x) = \begin{cases}
    17-12\ln(x/2) & x<1,\\
    17+12\ln(2) & x\geq1.
  \end{cases}
\end{equation}
To proceed with our estimates, we need the positron density in the
nugget's electrosphere at temperature $T$. As shown in
Appendix~\ref{sec:struct-nugg-surf}, the corresponding expression in
the nonrelativistic mean-field approximation is given by
\begin{equation}
  \label{eq:n_r}
  n(z,T)\simeq \frac{T}{2\pi\alpha}\cdot \frac{1}{(z+\bar{z})^2},
\end{equation}
where $\bar{z}$ is a constant of integration to be determined by some
appropriate boundary condition.  It is known that the mean-field
approximation is not valid for extremely large $z$, where exponential
rather than power-law (\ref{eq:n_r}) decay is expected.  We could
accommodate the corresponding feature by introducing a cutoff at
sufficiently large $z=z_{\text{max}}$.  The result, however, is not
sensitive to this cutoff, so we shall simply take
$z_{\text{max}}=\infty$ below to obtain our order of magnitude
estimate.

Note: the electrosphere extends well beyond the core of the nugget.
To see this, note that the Boltzmann regime~(\ref{eq:n_r}) is based on
an approximation that neglects the curvature of the nuggets surface
(see appendix~\ref{sec:struct-nugg-surf}).  This regime terminates
only once the curvature becomes significant, i.e. once the
electrosphere extends at least to the same order as the macroscopic
radius of the nugget.  Thus, the electrosphere occupies a significant
fraction of the nugget's volume: it is not just a thin outer shell.
Corrections from the finite size of the nugget are discussed in
appendix~\ref{sec:finite-size} but do not affect the magnitude
of our calculations.

The parameter $\bar{z}$ is not a free parameter, but is fixed by
matching the full density profile to the boundary of the nuclear
matter core of the nugget, where the lepton chemical potential
$\mu_{0}\approx 10\MeV$ is established by beta-equilibrium in the
nuclear matter.  A proper computation of $\bar{z}$ thus requires
tracking the density through many orders of magnitude from the
ultrarelativistic down to the nonrelativistic regime, which is
beyond the scope of this work.  The order of magnitude, however, is
easily estimated by taking $z=0$ as the onset of the Boltzmann regime:
\begin{subequations}
  \label{eq:zbar}
  \begin{align}
    n_{z=0} &= \frac{T}{2\pi\alpha}\cdot \frac{1}{\bar{z}^2} \simeq
    (mT)^{3/2},\\
    \bar{z}^{-1}
    &\simeq \sqrt{2\pi\alpha }\cdot m\cdot   \sqrt[4]{\frac{T}{m}}.
  \end{align}
\end{subequations}
Numerically, $\bar{z}\sim 0.5\cdot 10^{-8}$~cm while the density
$n\sim 0.3 \cdot 10^{23} $~cm$^{-3}$ for $T\simeq 1\eV$.  (See
Appendix~\ref{sec:struct-nugg-surf} for details of this calculation.)

Our next task is to estimate the surface emissivity (radiant exitance)
$F = \int\d{z}\;Q(z)$---defined as the energy $E$ emitted per unit
time $\d{t}$, per unit area $\d{A}$ (flux)---from the nugget's surface by
integrating the emissivity (\ref{eq:spectrum}) over the Boltzmann regime
$z\in[0,z_{\text{max}}\rightarrow\infty]$, and introducing an extra
factor $1/2$ to account for the fact that only the photons emitted
away from the core can actually leave the system.

Our final estimate for spectral surface emissivity can be
expressed as follows:
\begin{multline}
  \label{eq:P}
  \frac{\d{F}}{\d{\omega}}(\omega) = 
  \frac{\d{E}}{\d{t}\;\d{A}\;\d{\omega}}
  \simeq
  \frac{1}{2}\int^{\infty}_{0}\!\!\!\!\!\d{z}\;
  \frac{\d{Q}}{\d{\omega}}(\omega, z)
  \sim \\
  \sim
  \frac{4}{45}
  \frac{T^3\alpha^{5/2}}{\pi}\sqrt[4]{\frac{T}{m}}
  \left(1+\frac{\omega}{T}\right)e^{-\omega/T}h\left(\frac{\omega}{T}\right).
\end{multline}
Integrating over $\omega$ contributes a factor of
$T\int\d{x}\;(1+x)\exp(-x)h(x)\approx 60\,T$, giving the total surface
emissivity:
\begin{equation}
  \label{eq:P_t}
  F_{\text{tot}} = 
  \frac{\d{E}}{\d{t}\;\d{A}} = 
  \int^{\infty}_0\!\!\!\!\!\d{\omega}\;
  \frac{\d{F}}{\d{\omega}}(\omega) 
  \sim
  \frac{16}{3}
  \frac{T^4\alpha^{5/2}}{\pi}\sqrt[4]{\frac{T}{m}}\\
\end{equation}
Although a discussion of black-body radiation is inappropriate for
these nuggets (for one thing, they are too small to establish thermal
equilibrium with low-energy photons), it is still instructive to
compare the form of this surface emissivity with that of black-body
radiation $F_{BB}=\sigma T^4$:
\begin{equation}
  \label{eq:BB}
  \frac{F_{\text{tot}}}{F_{BB}} \simeq
  \frac{320}{\pi^3}\alpha^{5/2}\sqrt[4]{\frac{T}{m}}.
\end{equation}
At $T=1\eV$, the emissivity $F_{\text{tot}} \sim 10^{-6} F_{BB}$ is much smaller
than that for black-body radiation.  The spectral properties of these
two emissions are also very different at low frequencies $\omega\leq
T$ as follows from (\ref{eq:P}).

These two differences are essential to explain the \WMAP\ haze.  First,
the suppressed total radiant exitance is required to establish the eV
temperature scale (this will be discussed in
Section~\ref{sec:therm-nugg}).  Second, the extremely long
low-frequency tail due to the logarithmic dependence of $h(x)$ is
required to ensure that sufficient power is radiated in the microwave.
Thus, as we shall show in~\ref{sec:comp-wmap-haze}, it is highly
nontrivial that the scale of the emitted microwave emission should be
consistent with the observed \WMAP\ haze emission: If the nuggets had
been simple black-body emitters, the emission would be many orders of
magnitude below the observed scale.

Finally, we emphasize here that there are no free parameters in this
calculation: all of the scales are set by well-established nuclear and
electromagnetic physics.  The only unknown parameters that enter are the
overall normalizations which can be fixed by considering the related
diffuse x-ray emission.  This is the point of the next section.
\subsection{Thermodynamic Equilibrium}
\label{sec:therm-nugg}
Armed with an estimate of the total emissivity~(\ref{eq:P_t}), we may
discuss the thermodynamic properties of the nuggets.  In order to
maintain the overall energy balance, the nuggets must emit energy at
the same rate that it is deposited through proton annihilation,
\begin{equation}
  \label{eq:FtotFann}
  F_{\text{tot}} = (1-g)F_{\text{ann}}
  = (1-g)\frac{\d{E_{\text{ann}}}}{\d{t}\,\d{A}},
\end{equation}
where $1-g$ is the fraction of the annihilation energy that is
thermalized.  Note that both the rate of emission and the rate of
annihilation are expressed per unit area $A$, so that the equilibrium
condition is independent of the nugget size.  The rate of annihilation
$F_{\text{ann}}$ is
\begin{equation}
  \label{eq:Fann1}
  F_{\text{ann}} = 2\GeV\cdot f\cdot v \cdot n_{\VM}(\vect{r})
\end{equation}
where $2\GeV = 2\,m_{p}$ is the energy liberated by proton annihilation,
$v$ is the speed of the nugget through the visible matter,
$n_{\VM}(\vect{r})$ is the local visible matter density, and
\begin{equation*}
  f = \frac{\sigma_{\text{ann}}}{A} \sim 10^{-1}
\end{equation*}
is the factor by which the effective cross-section
$\sigma_{\text{ann}}$ for proton annihilation is reduced from the
geometric cross-section $A$ due to the possibility of reflection from
the sharp quark-matter surface (in contrast, the positron distribution
in the electrosphere is very smooth), as discussed in~\cite{Forbes:2006ba}.

The typical galactic scale for the speed is $v\sim 100 \text{ km/s}
\sim 10^{-3} c$, while the density at a distance $r\sim$ kpc from the
center is
\begin{equation*}
  n_{\VM} \sim \xi \cdot n_{\VM}^{\text{disk}} = 
  \xi \cdot\frac{3}{\text{cm}^{3}} \approx \frac{150}{\text{cm}^3}
\end{equation*}
where we have adopt a scaling behaviour close to that of an
isothermal sphere~\cite{Padmanabhan:2001} for the typical visible
matter density in the bulge at a distance $r\sim$ kpc from the core,
where the observed \WMAP\ haze originates:
\begin{equation*}
  \xi \approx \left(\frac{8.5\text{ kpc}}{r}\right)^{1.8} \sim 50.
\end{equation*}
Combining these, we obtain
\begin{equation}
  \label{eq:Fann2}
  F_{\text{ann}} \sim \frac{10^{9}\text{GeV}}{\text{cm}^2\cdot\text{s}}
  \cdot \left(\frac{f}{10^{-1}}\right) \cdot \left(\frac{v}{10^{-3}c}\right)\cdot
  \left(\frac{n_{\VM}}{300/\text{cm}^{3}}\right)
\end{equation}
which must be compared with the total surface emissivity~(\ref{eq:P_t})
\begin{equation*}
  F_{\text{tot}} \sim 10^{9}
  \frac{\text{GeV}}{\text{cm}^2\cdot\text{s}}
  \left(\frac{T}{\text{
eV}}\right)^{4+1/4}.
\end{equation*}
Taking the typical values $v \sim 10^{-3}c$ and $n_{\VM} \sim
300/$~cm$^3$ gives the relationship between the temperature and
typical parameters describing the nuggets, $f, g$,
\begin{equation}
  \label{eq:Equilibrium}
  \left(\frac{T}{\text{eV}}\right)^{4+1/4}
  \approx
  (1-g) \left(\frac{f}{10^{-1}}\right).
\end{equation}
As discussed in~\cite{Forbes:2006ba}, reasonable values for $f\sim
1/15$ and $g\sim 1/10$ all lead to a $T\sim\eV$ equilibrium
temperature.

The heat-capacity of the nuggets is estimated in
Appendix~\ref{sec:heat-capacities}.  If the gas of positrons occupies
a substantial volume of the nugget, then the heat capacity is
``large'' in the sense that it will require many annihilations to
raise the temperature of the nuggets to the eV scale at which
equilibrium is established.  Thus, the antimatter nuggets will act as
effective thermal integrators, slowly reaching a relatively constant
average temperature $T\sim$ eV.

\section{Explaining the WMAP Haze}
\label{sec:comp-wmap-haze}
In our proposal, interstellar matter annihilates on antimatter
nuggets.  The nuggets then radiate this energy over a wide range of
frequencies.  The model thus makes definite predictions relating these
emissions: they should have similar morphologies, and the relative
intensities should be related by an overall energy budget determined
by the local annihilation rate.

Four types of emission are from ``hot spots'' at the annihilation
sites, and should be observable from the core of our Galaxy:
\begin{myenumerate}[B.]
\item\label{item:511} Electron annihilations through positronium
  produce a well defined $511\keV$
  emission~\cite{Oaknin:2004mn,Zhitnitsky:2006tu} that is consistent
  with, and could possibly explain the puzzling diffuse $511\keV$
  emission observed by \SPI/\INTEGRAL.
\item\label{item:MeV} Direct electron annihilation can also
  produce emission in the $1$--$20\MeV$ band~\cite{Lawson:2007kp} which
  is consistent with, and could explain part of the diffuse
  gamma-ray emissions observed by \COMPTEL.
\item\label{item:x-ray} Proton annihilation produces keV x-ray
  emission from a hot spot at the annihilation
  site~\cite{Forbes:2006ba} that is consistent with, and could possibly
  explain the puzzling diffuse x-ray emissions observed by \Chandra.
\item\label{item:GeV} Proton annihilation occasionally produces GeV
  photons~\cite{Forbes:2006ba} that are consistent with, and could
  partially account for the gamma-ray emissions observed by \EGRET.
\end{myenumerate}
All of these emissions are ``direct'' in the sense that the timescale
for the emission is much shorter than the time between successive
annihilations.  Thus, the intensity of these emissions depends
only\footnote{There additional small dependencies that we neglect
  here: for example, on the local speed of the nuggets.  This
  introduces only small uncertainties, however, and certainly do not
  affect the overall magnitude.} on the rate of annihilation events
which is proportional to $n_{\VM}(r)n_{\DM}(r)$---the product of the
local visible and dark-matter distributions at the annihilation site.
The emitted spectrum is also independent of the local density.  We
emphasize that the model makes two nontrivial predictions:
1) that the morphology of these emissions is very strongly correlated,
and 2) that the spectral properties of these emissions are
\emph{independent of position}.

A comparison between observations of the ``direct'' emissions
B.\ref{item:511} through B.\ref{item:GeV} along the same line-of-sight
is possible because the local emission depends only on the
local rate of annihilation $\phi(\vect{r}) \propto
n_{\VM}(\vect{r})$.  The observed flux thus depends on the uncertain
matter distribution through the same line-of-sight integral:
\begin{equation}
  \label{eq:direct_integral}
  \Phi_{\text{511,x-ray,etc.}} \propto 
  \int\d{\Omega}\d{l}\, n_{\VM}(l)n_{\DM}(l),
\end{equation}
which cancels when comparing emissions from the same position in the
sky.

There is an additional emission from the nuggets:
\begin{myenumerate}[B.]
  \setcounter{myenumi}{4}
\item\label{item:WMAP} Energy not directly released through one
  of the mechanisms~B.\ref{item:511}--B.\ref{item:GeV} heats the nuggets,
  ultimately being thermally radiated.
\end{myenumerate}
As we have shown in Section~\ref{sec:therm-nugg} and
Appendix~\ref{sec:heat-capacities}, the heat-capacity and energy
budget ensure that the nuggets have a well-defined temperature scale
of $T\sim 1\eV$ in the core of the Galaxy.  The resulting thermal
bremsstrahlung emission thus ``averages'' the annihilations over time,
and the resulting emissivity and spectrum will depend on the
temperature $T(n_{\VM})$ which is a function of the local visible matter
density.

The observed flux of this thermal emission exhibits a slightly
different dependence because the local emission depends on the
temperature $\phi(\vect{r}) = \phi\{T[n_{\VM}(\vect{r})]\}$. As we shall
show Appendix~\ref{sec:line-sight-integr}, however, the difference is
small, and can be ignored for the order of magnitude comparisons we
present here.

Finally, in principle, we may compare the \emph{total} thermal
emission~(\ref{eq:P_t}) with the ``direct'' emissions because thermal
equilibrium relates the rate of total emission to the rate of
annihilation, both of which are proportional to $n_{\VM}(\vect{r})$.  In
practise, however, the thermal eV scale emission cannot be seen
against the bright stellar background.

\subsection{Energy Budget}
\label{sec:energy-budget}
In~\cite{Forbes:2006ba}, the ``direct'' emissions B.\ref{item:511},
B.\ref{item:x-ray}, and B.\ref{item:GeV} were compared, showing that
our proposal is consistent with the current observations, and using
the observations to constrain the properties of the nuggets.  In
particular, two parameters were introduced describe complicated
properties of the nuggets: The parameter $f<1$ was introduced to
describe the suppression of the proton annihilation rate with respect
to the electron annihilation rate, and the parameter $g <1/2$ was
introduced to describe the fraction of the proton annihilation energy
that is directly released as x-rays.  We emphasize that these
parameters are not free, but they depend on detailed models of the
nuggets and are beyond the reach of present day calculational
techniques.

By hypothesizing that emissions from the nuggets completely explain
the puzzling 511 keV (B.\ref{item:511}) and diffuse x-ray
(B.\ref{item:x-ray}) emissions, one obtains $fg\sim 6 \cdot 10^{-3}$,
which is consistent with the theoretical estimates, and provides a
nontrivial test of the theory.

Neglecting the small corrections to the line-of-sight averaging
discussed above, we may perform a similar analysis of the \WMAP\ haze to
see if it is also consistent with the our proposal.  The thermal
energy input into the antimatter nuggets is the complementary
fraction\footnote{Technically, we should include the energy from
  electron annihilations, and subtract the fraction
  $\alpha/\alpha_{s}$ of GeV photons~B.\ref{item:GeV}
  \cite{Forbes:2006ba}.  These are only small corrections to the
  overall energy budget.} $1-g$ of the total proton annihilation
energy not directly released as x-rays.  Thus, if we use the observed
x-ray flux $\Phi_{\Chandra}$ to provide the energy normalization, then
the total thermal emission will be approximately
\begin{equation}
  \label{eq:phi_T}
  \Phi_{T} \approx \frac{1-g}{g}\Phi_{\Chandra}.
\end{equation}
The total thermal emission $\Phi_{T}$ may then be used to estimate the
observed microwave emission in a specified frequency band by computing
the ratio $\gamma$ of spectral emissivity~(\ref{eq:P}) in the
specified band to the total emissivity~(\ref{eq:P_t}):
\begin{equation*}
  \gamma = \frac{1}{F_{\text{tot}}}
  \int_{\omega}^{\omega + \Delta\omega}\!\!\!\!\!\!\!
  \d{\omega}\;
  \frac{\d{F}}{\d{\omega}}(\omega)
  \approx 
  \frac{25 - 12\ln(\omega/T_{\text{eff}})}{60\,T_{\text{eff}}}\Delta\omega,
\end{equation*}
where $T_{\text{eff}}$ is an ``average'' temperature that accounts for
variations along the line of sight.  This is the fraction of the total
emitted thermal radiation emitted in the microwave band $\omega$ of
width $\Delta\omega\ll\omega\ll T_{\text{eff}}$.  For the typical
scale of the emissions we are considering, $T_{\text{eff}}\sim \eV$
and $\omega \sim h\cdot 30\GHz \sim 10^{-4}\eV$, so we have
\begin{equation*}
  \gamma \approx 2\frac{\Delta\omega}{T_{\text{eff}}}.
\end{equation*}
The total observed microwave flux is then related to the total thermal
flux (\ref{eq:phi_T}) by
\begin{equation*}
  \Delta\omega\frac{\d\Phi_{\WMAP}}{\d\omega} \approx 
  \gamma\Phi_{T} \approx \gamma \frac{1-g}{g}\Phi_{\Chandra},
\end{equation*}
giving us the relationship
\begin{equation}
  \frac{\d\Phi_{\WMAP}}{\d\omega} \approx 
  \frac{2}{T_{\text{eff}}}\frac{1-g}{g}\Phi_{\Chandra}.
\end{equation}
Observationally, \Chandra\ observes a total flux~\cite{Muno:2004bs}
\begin{subequations}
  \begin{equation}
    \Phi_{\Chandra} \approx 2\times
    10^{-6}\frac{\text{erg}}{\text{cm}^2\cdot\text{s}\cdot\text{sr}}
  \end{equation}
  while the \WMAP\ haze flux is
  \cite{Finkbeiner:2003im,Finkbeiner:2004je,Finkbeiner:2004us,Hooper:2007gi}
  \begin{equation}
    \frac{\d\Phi_{\WMAP}}{\d{\omega}} = (3\text{--}6)
    \frac{\text{kJy}}{\text{sr}} 
    \approx 
    \frac{(3\text{--}6)\times10^{-20}\text{erg}}
    {\text{cm}^2\cdot\text{s}\cdot\text{sr}\cdot\text{Hz}}.
  \end{equation}
\end{subequations}
Combining these, and converting $1$ Hz $\approx 4\times 10^{-15}\eV$,
we predict that the observed \WMAP\ haze intensity will be saturated by
thermal antinugget emission if the parameters which enter in our
estimate satisfy the following constraint
\begin{equation}
  \label{eq:chandra-wmap}
  \frac{\text{eV}}{T_{\text{eff}}}\cdot\frac{1-g}{g} \approx
  (2 \text{--} 4).
\end{equation}
Although this relationship is only approximate, it is quite amazing
that it is satisfied (without any adjustment) if the previous
estimates for $T_{\text{eff}}$ and $g$ are used.  Thus, the nontrivial
relationship~(\ref{eq:chandra-wmap})---which depends strongly on the
observed intensity of the GHz \WMAP\ haze---is satisfied by the
phenomenological parameters determined by considering only the keV
scale emissions.
\subsection{Spectrum}
\label{sec:spectrum}
The observed spectrum of the \WMAP\ haze is extremely
hard~\cite{Finkbeiner:2003im,Hooper:2007kb}.  This feature is easily
accommodated in our model by the logarithmic dependence of the thermal
bremsstrahlung emission~(\ref{eq:P}).  Indeed, the \WMAP\ haze was
initially interpreted as thermal bremsstrahlung (free-free emission)
from a hot ($T\sim$ eV) gas~\cite{Finkbeiner:2003im} ($10^4$~K$< T <
10^6$~K), but this interpretation was rejected because a $H\alpha$
recombination line, which should accompany the haze, is not seen. (The
possibility of much hotter plasmas $T\gg 10^4$~K has also been ruled
out~\cite{Finkbeiner:2003im,Finkbeiner:2004je,Finkbeiner:2004us,Hooper:2007kb}.)

It is quite remarkable that the $T\sim 10^4$~K$ \sim \eV$ scale arises
natural in our proposal in two completely independent ways:
(\ref{eq:Equilibrium}) and (\ref{eq:chandra-wmap}).  Our proposal,
thus, naturally fits the observed spectrum of the \WMAP\ haze, without
any $H\alpha$ recombination line since the emission is from a purely
positronic gas.  It is also remarkable that the spectrum exactly
corresponds with bremsstrahlung radiation, as was originally suggested
in \cite{Finkbeiner:2003im} to fit the data.

\subsection{Morphology}
\label{sec:morphology}
Our proposal also makes a definite prediction about the morphologies
of the various emissions.  In particular, the morphology of the
``direct'' emissions B.~\ref{item:511}--B.~\ref{item:GeV} should be
almost identical.  As discussed in
Appendix~\ref{sec:line-sight-integr}, even though the \WMAP\ haze is a
thermal emission, the dependence on the line-of-sight matter
distribution is quite weak, and our model thus also predicts that the
morphology of the \WMAP\ haze be closely correlated with the morphology of
the ``direct'' emissions.

Ultimately, if our proposal is correct, the morphology of all of the
emissions are direct probes of the matter distributions in our Galaxy,
and may thus become a useful measuring tool.

\section{Summary}
\label{sec:summary}
We have now demonstrated that our proposal naturally and nontrivially
explains diffuse galactic
radiations~B.\ref{item:511}--B.~\ref{item:WMAP} over thirteen orders of
magnitude from microwave ($10^{-4}$~eV) to GeV scales.  The only
``parameter'' in our model is the overall size of the nuggets, and the
dependence of the observations on this parameter can virtually be
eliminated by comparing observations along the similar lines-of-sight
to the core of the Galaxy.  Comparisons along the same line-of-sight
also virtually eliminate uncertainties related to the distribution of
matter in our Galaxy.  Removing this uncertainty, our proposal depends
on only a couple of parameters: $f<1$ and $g<1/2$.  These represent
presently intractable calculations, but have tight upper bounds and
can vary by no more than an order of magnitude or so.\footnote{The
  parameters $f$ and $g$ can, in principle, be calculated from the
  first principles.  However, such a computation is extremely
  difficult as it requires solving the many-body problem in a strongly
  coupled regime.  Indeed, the phase diagram of quark matter in the
  relevant regime is still largely unknown.  Even with these
  reservations, we still are quite confident that these parameters
  cannot deviate much from their ``natural'' values.  In this respect
  our proposal is predictive: there is little freedom to change these
  parameters.  This is in contrast with most other dark matter
  proposals where parameters can be arbitrary changed by
  many orders of magnitude to exploring an enormously large, and
  largely unconstrained parameter space.}

Together with our previous work~\cite{Forbes:2006ba}, we now have
several constraints on these parameters by postulating that matter
annihilation on dark-antimatter nuggets explains significant
proportions of puzzling diffuse emissions.  These
constraints provide a highly nontrivial test of our proposal.
\begin{myenumerate}[S.]
\item
  The diffuse 511 keV emission~B.\ref{item:511} observed by
  \SPI/\INTEGRAL\ has been identified as primarily due to positronium
  annihilation.  The puzzle is how positrons come to be diffusely
  distributed through the core of our Galaxy.  We propose that the
  positron electrosphere of dark-antimatter nuggets provide the source
  of positrons.  This source is distributed diffusely, and produces a
  definite spectrum consistent with the observed
  spectrum~\cite{Oaknin:2004mn,Zhitnitsky:2006tu} that is independent
  of any model-specific parameters.  One prediction is that the
  spectrum is independent of observed direction.  Another is that the
  intensity is determined by the product
  $n_{\VM}(\vect{r})n_{\DM}(\vect{r})$ of the distribution visible and
  dark matter.  We use this emission as a baseline to which we compare
  the other emissions in order to remove the uncertainties of the
  nugget size and the matter distribution along the line of sight to
  the galactic core.
\item Associated with this emission is the MeV spectrum from direct
  e$^{+}$e$^{-}$ annihilation~B.\ref{item:MeV}.  This spectrum is
  model independent, and consistent with observations and background
  models, possibly explaining the 1--20 MeV energy deficits seen in the
  \COMPTEL\ data \cite{Lawson:2007kp}.
\item The diffuse keV x-ray emissions~B.\ref{item:x-ray} measured by
  \Chandra\ are puzzling because of the implied energy
  budget.  The spectrum looks like a thermal $8\keV$
  plasma~\cite{Muno:2004bs}, but such a plasma would not even be
  gravitationally bound and would require a huge unidentified source
  of energy to fuel.  We propose that this emission is due to
  bremsstrahlung emission from positrons excited from protons
  annihilating on the dark-antimatter nuggets~\cite{Forbes:2006ba}.
  The spectrum for this process is also largely independent of
  model-specific parameters, is consistent with the observations, and
  is also independent of position.  Comparing this emission the 511
  keV emission gives one constraint on the parameters
  $fg$~\cite{Forbes:2006ba},
  \begin{equation}
    fg \approx 6\times10^{-3},
  \end{equation}
  that is satisfied by their natural scales $f\approx 1/15$ and
  $g\approx 1/10$.  The morphologies should also be related, depending
  on the product $n_{\VM}(\vect{r})n_{\DM}(\vect{r})$.
\item The direct emission of GeV photons from proton
  annihilation~B.\ref{item:GeV} is consistent with gamma ray
  observations by \EGRET, explaining up to one tenth or so of the
  observed spectrum~\cite{Forbes:2006ba}.
\item The annihilation energy not released through one of the previous
  sources of emission ultimately thermalizes.  As shown in
  Section~\ref{sec:therm-nugg}, the nuggets reach equilibrium with a
  typical $T\sim\eV$ scale through the
  constraint~(\ref{eq:Equilibrium}),
  \begin{equation}
    \tag{\ref{eq:Equilibrium}}
    \left(\frac{T}{\text{eV}}\right)^{4+1/4}
    \approx
    (1-g) \left(\frac{f}{10^{-1}}\right),
  \end{equation}
  which is again satisfied by the natural parameter scales.  We
  emphasize that this constraint is virtually model independent,
  depending on only the emissivity calculated in
  Section~\ref{sec:therm-emiss-from} and the properties of the matter
  distribution in the core of the Galaxy.  The emissivity is dominated
  by the properties of the nugget electrosphere in the low-density
  regime, where calculations are under order-of-magnitude control.
  Although not particularly sensitive to the emissivity, the scale set
  by~(\ref{eq:Equilibrium}) would be altered by an order of magnitude
  if the emissivity were black-body.
\end{myenumerate}

\subsection{Predictions}
\label{sec:predictions}
The constraints described in Section~\ref{sec:summary} did not involve
any measurements of the \WMAP\ haze.  Thus, the temperature $T\simeq\eV$~(\ref{eq:Equilibrium})
allows us to unambiguously predict the energy emitted in the tail of the
thermal distribution in the microwave band.  Amazingly, comparing this
with the observed \WMAP\ haze gives (\ref{eq:chandra-wmap})
\begin{equation}
  \tag{\ref{eq:chandra-wmap}}
  \frac{\text{eV}}{T_{\text{eff}}}\cdot\frac{1-g}{g} \approx (2\text{--}4),
\end{equation}
which is satisfied with the natural values of the parameters, even
though the frequencies are at a scale many orders of magnitude smaller
than the scale at which the parameters were constrained.

Unlike~(\ref{eq:Equilibrium}), this estimate is extremely sensitive to
the flat spectral properties of the thermal emission~(\ref{eq:P}) at
small frequencies $\omega \ll T$: a very specific feature of thermal
bremsstrahlung emission.  If the spectrum were not flat, there is no
way that this constraint would be satisfied.

Another difference between the two estimates~(\ref{eq:Equilibrium})
and (\ref{eq:chandra-wmap}) is that the former is sensitive overall
normalization of (\ref{eq:P_t}) whereas the latter is sensitive to the
detailed shape of the spectrum~(\ref{eq:P}).  We also emphasize that
the constraints (\ref{eq:Equilibrium}) and (\ref{eq:chandra-wmap}) not
only deal with emissions from scales separated by 8 orders of
magnitude, but that condition (\ref{eq:chandra-wmap}) depends on
observed intensity of the \WMAP\ haze which has absolutely nothing to do
with the estimate~(\ref{eq:Equilibrium}).  That these two estimates
agree with each other and with the estimates in~\cite{Forbes:2006ba},
(where the intensity of the diffuse keV x-rays were compared with the
511 keV emission), is truly remarkable.

In addition to satisfying the constraints described in the previous
section, our proposal also makes the definite prediction that
the morphologies of the 511\keV\ flux, the 1--20 MeV $\gamma$-ray
emission, and the x-ray flux should all be identical, following the
distribution $n_{\VM}(\vect{r})n_{\DM}(\vect{r})$ of visible and dark
matter.  The morphology of the \WMAP\ should be similar but may differ
slightly because of the different line-of-sight
integral~(\ref{eq:thermal_integral}).  For example, \Chandra\ has
detected a diffuse x-ray emission with flux $6.5\times
10^{-11}$~erg/cm$^{2}$/s/deg$^{2}$ from a region of the disk
$28^\circ$ off the center~\cite{Ebisawa:2005zt}.  This is one order of
magnitude smaller than the observations from the core of the Galaxy,
and so our model predicts that the microwave emission from this region
is about one order of magnitude smaller than the \WMAP\ haze from the
galactic core, which seems consistent with the
observations~\cite{Finkbeiner:2003im}.

Finally, this proposal makes definite testable predictions for the
properties of the emitted spectra.  In particular, all bands but the
\WMAP\ haze are produced on an event-by-event basis, and are thus
independent of the rate at which the annihilation processes occur.
The observed spectra should thus be largely \emph{independent of the
  direction of observation.}  Only the intensity should vary as a
function of the collision rate, and this should be correlated with the
visible/dark-matter distribution as discussed above.  The \WMAP\ haze
will have a slight spectral dependence, but only through the temperature
dependence $T(n_{\VM})\propto n_{\VM}^{4/17}$ which is quite weak.
\subsection{Conclusion}
\label{sec:conclusion}
Our dark matter proposal not only explains many astrophysical and
cosmological puzzles, but makes definite predictions about the
correlations of the dark and visible matter distributions
$n_{\VM}n_{\DM}$ with five different bands of radiation ranging over
13 orders of magnitude in frequency.  In addition, it makes the
definite prediction that these spectra of the emissions should be
virtually independent of the local environment.  Such correlations and
spectral properties would be very difficult to account for with other
dark-matter candidates.  Future observations may thus easily confirm
or rule out this theory.  If confirmed, it would provide a key for
many cosmological and astrophysical secrets, and finally unlock the
nature of dark matter.

\acknowledgments 

The authors would like to thank to members of the UBC astronomy group,
especially M.~Halpern, H.B.~Richer, and D.~Scott, for providing a
forum for vetting the ideas presented in this paper, and for useful
discussions.  This research was supported in part by the US Department
of Energy under Grant No. DE-FG02-97ER41014, and the Natural
Sciences and Engineering Research Council of Canada.
\ifx\mcitethebibliography\mciteundefinedmacro
\PackageError{apsrevM.bst}{mciteplus.sty has not been loaded}
{This bibstyle requires the use of the mciteplus package.}\fi

\appendix
\section*{Appendix: Detailed Calculations}
\label{sec:deta-calc}
\subsection{Structure of Electrosphere}
\label{sec:struct-nugg-surf}
Here we briefly discuss the properties of the antimatter nugget
electrosphere to set the scales of the problem and determine the
structure required in Section~\ref{sec:therm-emiss-from}.

The radius of the nuggets depends on the mass, but must be larger than
$R > 10^{-7}$~cm at the lower limit $|B|>10^{20}$ set by terrestrial
nondetection.  We expect that the most likely size is on the order of
$|B| \sim 10^{25\text{-}27}$~\cite{Forbes/Zhitnitsky:2007b}.  The
quark-matter core of the nuggets ends sharply on a fm scale as set by
nuclear physics.  Near the surface, as the density falls, the quark
matter will definitely be charged due to the relatively large mass of
the strange quark $m_{s}\sim
100\MeV$~\cite{Alcock:1986hz,Kettner:1994zs,Madsen:2001fu,Usov:2004iz},
however, depending on the phase of quark matter realized in the core,
the matter may be charged throughout.  Charge neutrality will be
maintained through beta-equilibrium, which will establish a positron
chemical potential $\mu_{\text{e}^{+}} = \mu_{0} \simeq 10\MeV$.  (The
precise value depends on specific details of the quark-matter phase
and may range from a few\MeV\ to hundreds of\MeV, but is about an
order of magnitude less than the quark chemical potential
$\mu_{q}\simeq 500 \MeV$~\cite{Alcock:1986hz,Steiner:2002gx}.)  This
will induce a thin but macroscopic ``electrosphere'' of positrons
surrounding the quark-matter core in the transition region as
$\mu_{\text{e}^{+}}\rightarrow 0$ in the vacuum.

The structure of this electrosphere has been considered for quark
matter~\cite{Alcock:1986hz,Kettner:1994zs}, and the existence of this
``transition region'' is a very generic feature of these systems.  It
is the direct consequence of Maxwell's equations and chemical
equilibrium.  The region is called the electrosphere, emphasizing the
fact that quarks and other strongly interacting particles are not
present.  In the case of antimatter nuggets the ``electrosphere''
comprises positrons.

The variation of this chemical potential $\mu_{\text{e}^{+}}(z)$, and
the density $n(z)$ as a function of distance $z$ from the surface of
the nugget may be computed using a mean-field treatment of the
Maxwell equations~\cite{Alcock:1986hz,Kettner:1994zs,Usov:2004kj}.
For example, in the relativistic regime, one has~\cite{Usov:2004iz}
\begin{align}
  \label{eq:field}
  \mu_{\text{e}^{+}}(z)
  &=\sqrt{\frac{3\pi}{2\alpha}}\frac{1}{(z+z_{0})},\nonumber\\
  n(z) &\approx \frac{\mu^3_{\text{e}^{+}}}{3\pi^2} 
  = \frac{1}{3\pi^2}
  \left(\frac{3\pi}{2\alpha}\right)^{3/2}\frac{1}{(z+z_{0})^3},\\
  z_{0} &= \sqrt{\frac{3\pi}{2\alpha}}\frac{1}{\mu_0}, \nonumber
\end{align}
where $\mu_0\equiv \mu_{\text{e}^{+}}(z=0) \sim 10\MeV$ is the chemical
potential realized in the nugget's bulk.  The corresponding results
can be obtained outside of the relativistic regime, but they do not
have a simple closed form.  These calculations treat the electrosphere
as a one-dimensional wall rather than including the full radial
structure, essentially keeping only the first term in the $z/R$
expansion.  This approximation does not affect the order of magnitudes
of our calculation and will also be employed here.

The majority of the thermal emission considered in this work comes
from the nonrelativistic regime, which we may also analyze
analytically using the Boltzmann approximation.  The mean-field
approximation amounts to solving the Poisson equation
\begin{equation}
  \label{eq:Poisson_1}
  \nabla^{2}\phi(\vect{r}) = - 4\pi e n(\vect{r})
\end{equation}
where $\phi(\vect{r})$ is the electrostatic potential and
$n(\vect{r})$ is the density of positrons.  Using the spherical
symmetry of the nuggets and making the one-dimensional approximation,
we can write this as\footnote{Here we drop the radial term
  $2\phi'(z)/r$ on the left hand side of (\ref{eq:Poisson_1}),
  assuming that the radius of the nuggets $R\gg z$ is much larger than
  the thickness of the electrosphere.}
\begin{equation}
  \diff[2]{\phi(z)}{z} = - 4\pi e n(z)
\end{equation}
where $z$ is the distance from the quark nugget surface.  We now
introduce the positron chemical potential $\mu_{\text{e}^{+}}(z) =
-e\phi(z)$ which is the potential energy of a charge at position $z$
relative to $z=\infty$ where we take $\mu_{\text{e}^{+}}(\infty) = 0$
as a boundary condition.  This gives
\begin{equation}
  \label{eq:Poisson}
  \frac{\d^2\mu_{\text{e}^{+}}(z)}{\d{z}^2} 
  = 4\pi\alpha\, n[\mu_{\text{e}^{+}}(z)]
\end{equation}
with the additional boundary conditions $\mu_{\text{e}^{+}}(z=0) =
\mu_{0} \sim 10$ MeV as established by beta-equilibrium in the quark
matter.  Here $n[\mu_{\text{e}^{+}}]$ is the density of a free
Fermi-gas of positrons as a function of the chemical potential for the
positrons.  The full relativistic form is
\begin{equation}
  \label{eq:n_mu}
  n[\mu] = 2\!\int \frac{\d^{3}{p}}{(2\pi)^3}\; \left(
    \frac{1}{1+e^{\frac{\epsilon_{p}-\mu}{T}}}
    -
    \frac{1}{1+e^{\frac{\epsilon_{p}+\mu}{T}}}\right),
\end{equation}
where $\epsilon_{p} = \sqrt{p^2+m^2}$, which has the property
$n[\mu=0] = 0$ required by our identifying the chemical potential
$\mu$ with respect to $z=\infty$.  This is quite complicated, but is
well approximated in the nonrelativistic Boltzmann regime where $n
\ll (mT)^{3/2}$ by
\begin{multline}
  \label{eq:Boltzman_n}
  n[\tilde{\mu}] \approx 2\int \frac{\d^{3}{p}}{(2\pi)^3}\;
  e^{[\tilde{\mu} - p^2/(2m)]/T} =\\
  = \sqrt{2}\left(\frac{mT}{\pi}\right)^{3/2}e^{\tilde{\mu}/T}
\end{multline}
where we have performed a nonrelativistic expansion $\sqrt{p^2+m^2}
\approx m + p^2/(2m)$, dropped the antiparticle contribution, and
neglected the quantum degeneracy.  The effective chemical potential
$\tilde{\mu} = \mu_{\text{e}^{+}} - m$ is related to the vacuum chemical
potential $\mu$ by subtracting the mass.  We note that the right
boundary condition must now be changed to $n(z=\infty) = 0$ because
$\tilde{\mu}$ does not tend to zero under these approximations.  The
left boundary condition must be determined by matching the density at
some point to the full relativistic solution that integrates to the
quark-matter core and matches~(\ref{eq:field}).  The differential
equation~(\ref{eq:Poisson}) may now be expressed in terms of either
$\tilde{\mu}(z)$ or $n(z)$, leading to the peculiar solution
\begin{equation}
  \label{eq:n_z}
  n(z) = \frac{T}{2\pi\alpha}\frac{1}{(z+\bar{z})^2},
\end{equation}
where $\bar{z}$ is an integration constant determined by matching to a
full solution.  Here we make a simple approximation, defining $z=0$ as
the onset of the Boltzmann regime:
\begin{equation}
  n(z=0) = \frac{T}{2\pi\alpha\bar{z}^2} = (mT)^{3/2}.
\end{equation}
While not an exact matching procedure, this will give the correct
parametric dependence and will be valid for the order-of-magnitude
estimates required.  Thus, we have the following characterization of
the density in the Boltzmann regime as used in
Section~\ref{sec:therm-emiss-from}:
\begin{equation}
  \bar{z}^{-1} \simeq \sqrt{2\pi\alpha}\cdot m \cdot\sqrt[4]{\frac{T}{m}}.
\end{equation}
The region where $z<0$ here corresponds to the region of higher
density closer to the nugget's surface where the Boltzmann
approximation breaks down due to degeneracy effects. One can argue,
however, that in this degenerate regime, the emissivity is strongly
suppressed for two reasons: 1) only a small portion of the particles
close to the Fermi surface can participate in scattering processes, so
the phase space for emission is dimensionally reduced, and 2) as the
density increases, the plasma frequency increases, and low-energy
photons cannot escape.

Our estimates include only emission from the Boltzmann regime.  In
principle, the emission from denser regions could contribute at the
same order: this approximation thus underestimates the emission by a
factor, but not by an order of magnitude.

We make one final set of remark in response to the criticism
\cite{Cumberbatch:2006bj} (version 3) of our proposal.  As we have
shown, the density profile behaves very differently in different
physical regimes.  In the ultrarelativistic regime~(\ref{eq:field}),
one has the dependence $n\propto V^{3}$ where $V(z)$ is the
electrostatic potential in the mean-field
approximation~\cite{Alcock:1986hz,Kettner:1994zs,Usov:2004kj}, whereas
in the nonrelativistic Boltzmann regime~(\ref{eq:Boltzman_n}), one
has $n\propto e^{V/T}$, which is the well-known expression for the
density in nonrelativistic and nondegenerate
systems~\cite{LL5:1980}.  In the intermediate regimes, the dependence
is quite complicated due to the competing scales that appear in
$n[\mu]$~(\ref{eq:n_mu}).  Thus, one cannot simply apply formulae
like $n\propto V^{3}$ derived in one regime to describe physics in
another as was done in~\cite{Cumberbatch:2006bj}.

In general, one must also take into account quantum many-body
effects---such as charge screening (completely ignored
in~\cite{Cumberbatch:2006bj}), the plasma frequency, etc.  In the
Boltzmann regime discussed here, the density is sufficiently low that
many-body corrections may be neglected and vacuum results, such as the
cross-section~(\ref{eq:sigma}), employed.  At higher densities,
however, when the degeneracy becomes important---of order roughly $n
\geq (mT)^{3/2}$ in our case---many-body effects can drastically alter
the behaviour of the system and cannot be neglected as they were
in~\cite{Cumberbatch:2006bj}, even when considering only the
qualitative physics.

\subsection{Boltzmann Averages}
\label{sec:boltzmann-average}
To evaluate equation~(\ref{eq:spectrum}) we need to perform the
thermal average
\begin{equation}
  \left\langle\abs{v_{1} - v_{2}}\left(
      17 + 12\ln\frac{(p_{1} - p_{2})^2}{m\omega}
    \right)\right\rangle.
\end{equation}
As we are in the Boltzmann regime, we may simplify the calculation by
computing this in the Boltzmann ensemble.  Formally, we must integrate
over both momenta $p_{1}$ and $p_{2}$, but as we are only interested
in the order of magnitude, we simply perform the average over only a
single momentum $p_{2}$, setting $p_{1}=0$.\footnote{Including the
  full angular integrals increases the result by a factor of about
  $\sqrt{2}$ or so.
  \begin{multline*}
    \left\langle
      v_{12}\left(17+12\ln\frac{mv_{12}^2}{\omega}
      \right)
    \right\rangle = \\
    = 4\sqrt{\frac{T}{m\pi}}
    \left(1+\frac{\omega}{T}\right)e^{-\omega/T} \left(17 +
      12\tilde{g}\left(\frac{\omega}{T}\right) \right).
  \end{multline*}
  The calculation of $\tilde{g}(x)$, however, is somewhat tricky.}
\begin{multline}
  \left\langle
    v_{12}\left(17+12\ln\frac{mv_{12}^2}{\omega}\right)
  \right\rangle
  =\\
  =
  2\sqrt{\frac{2T}{m\pi}}
  \left(1+\frac{\omega}{T}\right)
  e^{-\omega/T}
  h\left(\frac{\omega}{T}\right)
\end{multline}
where
\begin{align*}
  h(x) &=  17 + 12g(x), \\
  g(x) &= \ln(2) + \frac{1+E_{1}(x)e^{x}}{1+x},\\
  E_{1}(x) &= \int_{1}^{\infty}\frac{e^{-xz}}{z}\d{z}.
\end{align*}
The following approximation for $h(x)$ is accurate to within $25\%$
for all $x$:
\begin{equation}
  h(x) = \begin{cases}
    17-12\ln(x/2) & x<1,\\
    17+12\ln(2) & x\geq1.
  \end{cases}
\end{equation}

\subsection{Finite-Size Effects}
\label{sec:finite-size}
So far, our calculations have assumed that we are working in infinite
matter.  Here we estimate the size of the corrections due to the fact
that the nuggets have a finite extent on the order of $L \geq
10^{-5}$~cm.  We shall demonstrate that properly accounting for these
corrections does not significantly affect our estimates of the
microwave band emission (though it may drastically suppress emission
at much longer wave lengths).

In principle, finite-size effects may change the cross section
(\ref{eq:sigma}), and therefore, our estimation of the emissivity
(\ref{eq:spectrum}).  The cross-section (\ref{eq:sigma}) was derived
using a continuum of plane-wave states, whereas to account for the
finite-size effects, one should use the basis of states bound to the
quark core.  To estimate the size of the corrections, one can imagine
confining the positrons to a box of finite extent $L$.

The electromagnetic field may still be quantized as in free-space with
states of arbitrarily large size because the photons are not bound to
the core, and are not in thermodynamic equilibrium with the positrons.
Their mean-free-path is much larger than $L$, so the low-energy
photons produced by the mechanism described above will simply leave
the system before they have a chance to interact with other positrons.

Therefore, it is only the positron states that must be considered over
a finite-size basis, which will modify the corresponding Green's
function used in the calculation of the cross-section
(\ref{eq:sigma}).  These modifications occur for momenta of the scale
$\delta p\sim \hbar/L$.  If $L \geq 10^{-5}$~cm, then this corresponds
to shifts in the energies of $\delta E \sim (\delta p)^2/2m \sim
10^{-6}$~eV~$\ll 10^{-4}$~eV, which is much smaller than the
transitions responsible for the emission at microwave frequencies.
Thus, we conclude that finite-size effects do not drastically change
the positron Green's function in the region of interests.  In
other-words, the expression for the cross-section
(\ref{eq:sigma})---derived using the standard (infinite volume)
Green's functions---remains valid for the estimation of the emission
and spectrum down to the microwave region $10^{-4}$~eV.  We also note
that finite-size effects do not change our estimates for the density
(\ref{eq:n_mu}, \ref{eq:Boltzman_n}) because the finite-size effects
$\delta{E} \ll T$ are much smaller than the typical energetic scale $T
\sim $~eV of the problem.  Thus, our expression (\ref{eq:sigma})
remains valid for photon energies $\omega \geq 10^{-4}$~eV.  To
calculate the emission of radiation with much longer wavelengths,
however, requires one to account for these finite-size corrections,
and we expect the emission of extremely low-energy photons $\omega \ll
\delta E \lesssim 10^{-6}$~eV to be suppressed.

\providecommand{\exclude}[1]{}
\exclude{
B:=1E22
L:=B^(1/3)/1e11*0.036mm
dp:=hbar/L
dE:=dp^2/2/m_e
dE -> eV
hbar^2/2/m_e*(0.036mm)^-2/(3e-4eV)
9.7993360253847502214e-8

*** Note as long as L > 1e-5 we are okay, but at 1e-6 we are in
trouble. This corresponds to B > 10^(24)*** }

One may ask how microwave radiation may be emitted from the nuggets
when the wavelength $\lambda$ is much larger than the size of the
nugget $\lambda\gg L$.\footnote{We thank a referee for raising this
  question.}  In general this is not a problem---consider the
well-known astrophysical emission of the $\lambda= 21$~cm line from
hydrogen with a size $a\simeq 10^{-8}$~cm---but there is a potential
suppression.  The coherence time $\tau$ of the positrons which must be
compared with the formation time $\sim\omega^{-1}$ of the photons.  If
the coherence time is too short, then multiple scatterings will
disrupt the formation of the photons.  This suppression is a case of
the so-called Landau-Pomeranchuk-Migdal (\textsc{lpm}) effect.

To estimate the coherence time $\tau$, consider the cross-section
$\sigma_{ee}$ of the positron-positron interaction.  This scales as
$\sigma_{ee}\sim \alpha^2/q^2$ where $q\sim b^{-1}$ is the typical
momentum transfer, and may be expressed in terms of the impact
parameter $b\sim n^{-1/3}$, which is estimated in terms of average
interparticle spacing where $n$ is the local positron density.

The mean-free-path $l$ is thus $l^{-1}\sim \sigma_{ee}n\sim \alpha^2n^{1/3}$.
Therefore, the typical time between collisions (which is the
same as coherence time) is $\tau\sim l/v$ where $v\sim
\sqrt{T/m}$ is the typical positron velocity.

Collecting all factors together we arrive at the estimate
\begin{equation}
  \omega\tau
  \sim \frac{\omega}{\alpha^2n^{1/3}}\sqrt{\frac{m}{T}}
  \sim \frac{\omega}{\alpha^2T}\left(
    1+\frac{z}{\bar{z}}\right)^{\frac{2}{3}}
  \geq 1.
\end{equation}
It is clear that this condition is satisfied for $\omega\geq
10^{-4}$~eV and $T\leq 1$~eV, even for $z=0$.  Thus, we were justified
in omitting \textsc{lpm} effect in our estimates in the low-density
regime (\ref{eq:n_z}).  However, from the same estimate it is clear
that this suppression becomes important for either smaller frequencies
$\omega \ll 10^{-4}$~eV or at higher densities.  We shall now show
that there us a much more significant high-density suppression than
the $\textsc{lpm}$ effect, which effectively turns off emission from
the bulk of the nuggets where the positron density is significantly
higher than (\ref{eq:n_z}).

\subsection{Emission from Very Dense Regions}
\label{sec:dense-regions}
Here we estimate the emissivity from very dense regions of the nugget
when the Boltzmann regime breaks down and degeneracy plays a
crucial role.  As we shall argue below, the corresponding emission from
very dense regions can be neglected in comparison with the estimates
(\ref{eq:P}) and (\ref{eq:P_t}) used in the text.

We start from (\ref{eq:spectrum}), which remains valid for any
densities $n(z, T)$.  To deal with denser regions, however, one must
the full expression for (\ref{eq:n_mu}) which includes the effects of
quantum degeneracy.  In these dense regions, only the states close to
the Fermi surface are excited and can participate in emission: the
states deep within the Fermi surface are ``Pauli blocked'' and cannot
participate in low-energy interactions.  It is the density of these
``quasiparticles''---not the full density---which enters the
emissivity calculations.  The other key property for these estimates
is the is the plasma frequency $\omega_p$, which characterizes the
propagation of photons in the degenerate systems.  For
ultrarelativistic systems, $\omega_p^2=\frac{4\alpha\mu^2}{3\pi}$
while for nonrelativistic systems, $\omega_p^2=\frac{4\pi\alpha
  n}{m}$.  The plasma frequency can be thought as an effective mass
for the photon: only photons with energy larger than this mass can
propagate outside of the system.  Photons with $\omega < \omega_{p}$ 
are ``off-shell'' or ``virtual'': these can only propagate for a short
period of time/distance $\sim \omega_{p}^{-1}$ before they decay.

To derive the analogues of equations (\ref{eq:P}) and (\ref{eq:P_t})
for the denser regions, we must start with (\ref{eq:spectrum}), but
insert the proper form for the expression for $n_1(z,T)n_2(z,T)$,
including these effects:
\begin{multline}
  \label{eq:n}
  n_1(z,T)n_2(z,T) \rightarrow \\
   4\!\int \frac{\d^{3}{p_1}}{(2\pi)^3} \frac{\d^{3}{p_2}}{(2\pi)^3}
    \frac{\theta(\tilde{\epsilon}_{p_1}+\tilde{\epsilon}_{p_2}
      - \omega)\theta(\omega-\omega_p)}
    {\left(1+e^{\frac{\epsilon_{p_1}-\mu(z)}{T}}\right)
      \left(1+e^{\frac{\epsilon_{p_2}-\mu(z)}{T}}\right)},
\end{multline}
where $\tilde{\epsilon}_{p_1}$ and $\tilde{\epsilon}_{p_2}$ are the
colliding positron quasiparticle energies above the Fermi
surface.  The factor of $\theta(\tilde{\epsilon}_{p_1} +
\tilde{\epsilon}_{p_2} - \omega)$ accounts for energy conservation:
the initial energy must be larger than the energy of emitted photon;
and the factor $\theta(\omega-\omega_p)$ accounts for the effects of
the plasma frequency.  Only photons with $\omega >\omega_p$ can
propagate in the dense media: photons with smaller energies will be
absorbed on distances $\sim \omega_{p}^{-1}$.  As we shall see, this
leads to an exponential suppression of emission $\sim \exp
(-\omega_{p}/T)$ when $\omega_p\gg T$.  As such, we have omitted
the aforementioned \textsc{lpm} suppression and additional Pauli
blocking effects, as these are comparatively insignificant.

The integral in (\ref{eq:n}) can be estimated from
\begin{equation*}
  \int\d\tilde{\epsilon}_{p_{1}}\d\tilde{\epsilon}_{p_{2}}
  \frac{\theta(\tilde{\epsilon}_{p_1}+\tilde{\epsilon}_{p_2}-\omega)
    \theta(\omega-\omega_p)}
  {\left(1+e^{\tilde{\epsilon}_{p_{1}}/T}\right)
    \left(1+e^{\tilde{\epsilon}_{p_{2}}/T}\right)}
  \sim T^2 e^{-\omega_{p}/T}
\end{equation*}
to give 
\begin{equation}
  \label{eq:n1}
  n_{1}(z,T)n_{2}(z,T) \sim 
  \frac{p_{F}^2\mu^2(z) T^2}{\pi^4}
  \exp{\left(-\frac{\omega_p(z) }{T}\right)}.
\end{equation}
The main point is that $\omega_{p}$ grows with the density as
$\omega_{p}\sim \sqrt{n}$ in the nonrelativistic regimes, and as
$\omega_{p}\sim \sqrt[3]{n}$ in the relativistic regimes.  This leads
to an exponential suppression of emission from the dense regions of
the nugget.  The suppression is lifted only when $\omega_{p}\sim T$
which occurs (as can be verified numerically) only when the densities
are sufficiently low that the Boltzmann approximation is valid.  Our
estimates (\ref{eq:P}) and (\ref{eq:P_t}), which are based on this
approximation, are thus justified.

One can estimate that the plasma frequency $\omega_{p}$ is a few eV
for densities (\ref{eq:n_z}) typical of the Boltzmann regime.  Given
our previous discussion, one might ask: How can low-energy photons
$\omega < \omega_{p}$ still be emitted?  The reason is that, although
these photons would be reabsorbed in infinite matter, this
reabsorption happens on a length scale of $\omega_p^{-1}$.  At the
typical densities in the Boltzmann regime, $\omega_{p}^{-1}\sim
0.3\cdot 10^{-5}$~cm is sufficiently large compared with the
nugget size that many of these photons will have left the nugget
before being reabsorbed.  One can interpret this effect as a decay of
a quasistationary state in quantum mechanics through the tunnelling
process, where the barrier has a size comparable with inverse energy
of the quasistationary state.  The suppression factor only becomes
sufficiently large when $\omega_{p} > L^{-1}$.  Then the reabsorption
happens well before the photons emerge.  This is the origin of the
primary suppression in the very dense regions we have just discussed.
\subsection{Heat Capacity}
\label{sec:heat-capacities}
Here we make a rough estimate of the heat capacity of the nuggets.  At
eV temperatures, the only modes that can be excited are the neutral
Nambu-Goldstone superfluid mode, which contributes as $T^{3}$, and the
gapless positrons at their Fermi surface of $p_{F} \approx 10\MeV$.
It is the gapless positrons that will dominate the low-temperature
heat capacity of the nuggets.

To estimate the heat capacity, we simply count the number of
low-energy modes.  The heat capacity density for a single mode with
dispersion $\epsilon_{p}$ is
\begin{equation}
  c_{V} = \int \frac{d^{3}p}{(2\pi)^3}\; \epsilon_{p} \frac{d}{dT}
  \left(\frac{1}{e^{\epsilon_{p}/T}\pm 1}\right).
\end{equation}
The total heat capacity $C_{V} \approx Vc_{V}$ is obtained by
integrating this over the volume $V$.  Here are some approximate
contributions to the heat capacity from the nuclear core, assuming the
density $\rho \approx 5 \GeV/$fm$^3$ is a few times nuclear density so
that $V\approx 26\, B \text{ GeV}^{-3}$, and taking $T\approx 1$ eV:
\begin{description}
\item[Single Boson:] For a single boson, $\epsilon_{p} = p/3$.
  This gives $c_{V} \approx 35\,T^{3}$ and a total contribution of
  $C_{V} \sim 10^{-24}B$.
\item[Two Fermions:] For a pair of positrons with Fermi surface $p_{F}$
  we have $\epsilon_{p} \approx v\abs{p-p_{F}}$ which gives $c_{V}
  \approx p_{F}^2 T/(3v)$ and a total contribution of $C_{V} \sim
  10^{-12}B$ where we take the Fermi velocity $v\approx c$ for
  relativistic systems and $p_{F}\approx 10$ MeV.
\end{description}
In the complete absence of positrons, the heat capacity of a
medium nugget $B\approx 10^{24}$ would be about unity, meaning that the
addition of $1\GeV$ of thermal energy would raise the temperature by a
GeV, however, even if the core of the nuggets is completely neutral
(for example, if the Colour-Flavour-Locked (\textsc{cfl}) phase were realized
with equal numbers of up, down, and strange quarks), the surface of
the quark-matter core will still be charged, requiring a large number
of positrons in the electrosphere to neutralize the objects.

Once one includes the contribution of the positrons, however, the heat
capacity becomes much larger, and the input of $1\GeV$ of energy will
only effect a very small change in temperature.  For example, if
positrons are present in the core, even the very smallest possible
nuggets\footnote{The constraint that the average $B>10^{21}$ is quite
  strong and comes from negative terrestrial based search
  results~\cite{Belolaptikov:1998mn}.} $B\sim 10^{21}$ would heat by
only an eV given $1\GeV$ of energy.  We expect the nuggets
have a typical size of $B\sim 10^{25\text{-}27}$
\cite{Forbes/Zhitnitsky:2007b}, so nuggets would require many
collisions in order to reach the eV temperature scales at which they
would thermally equilibrate by balancing the annihilation energy input
with the thermal radiation~(\ref{eq:Equilibrium}).

Note, however, that while the heat capacity is large in the sense that
many collisions are required to heat the nuggets to the equilibrium eV
scale, it is small enough to allow the nuggets to heat to this
temperature quickly on galactic scales.  To see this, consider the
total annihilation rate~(\ref{eq:Fann1}).
\begin{equation}
  \label{tau}
  P_{\text{ann}} = \frac{\d{E}}{\d{t}} = A\,F_{\text{ann}}
  \sim \left(\frac{B}{10^{25}}\right)^{2/3}\frac{\GeV}{\text{s}}
\end{equation}
where $A$ is the surface area of the nuggets
\begin{equation}
  A = 4\pi\left(\frac{3B\GeV}{4\pi\rho}\right)^{2/3}
  = 1.65 B^{2/3} \text{fm}^2.
\end{equation}
Therefore, the typical time between annihilations is on the order of
1 per second for nuggets with $B \sim 10^{25}$.  A nugget with
positrons throughout (corresponding to a larger heat capacity) would
require
\begin{equation}
  t_{\text{heat}} \sim \frac{1\eV\cdot C_{V}}{P_{\text{ann}}}
  \sim \left(\frac{B}{10^{25}}\right)^{1/3}\text{ hours}
\end{equation}
to reach eV temperatures.  Thus the nuggets reach their equilibrium
quite rapidly once they enter a region of high visible matter
density.
\subsection{Line-of-Sight Integration}
\label{sec:line-sight-integr}
A comparison between the ``direct''
emissions B.\ref{item:511}--B.\ref{item:GeV} is facilitated by the
fact that they depend on the same line-of-sight
average~(\ref{eq:direct_integral})
\begin{equation*}
  \Phi_{\text{511,x-ray,etc.}} \propto 
  \int\d{\Omega}\d{l}\, n_{\VM}(l)n_{\DM}(l).
\end{equation*}
In principle, comparing these with thermal emissions is more difficult
because the line-of-sight integral has an additional dependence on the
visible matter distribution.  This arises because the emissivity
depends on the temperature $T$ of the nuggets, which in turn depends
on the annihilation rate through the thermal equilibrium
condition~(\ref{eq:Equilibrium}).

As can be seen from~(\ref{eq:P_t}), (\ref{eq:FtotFann}) and
(\ref{eq:Fann1}), the temperature of the nuggets is related to the
local visible matter density as $T(n_{\VM}) \propto n^{4/17}_{\VM}$,
whereas from~(\ref{eq:P}), we see that the microwave emission scales
as $\phi(T) \propto T^{13/4} \propto n_{\VM}^{13/17} =
n_{\VM}n_{\VM}^{-4/17}$.  Thus, the microwave emission depends on the
line-of-sight integral
\begin{equation}
  \label{eq:thermal_integral}
  \Phi_{\WMAP} \propto 
  \int\d{\Omega}\d{l}\,\left[n_{\VM}(l)n_{\DM}(l)\right]n_{\VM}^{-4/17}(l).
\end{equation}
Of course, one must also account for rescattering, absorption and
other effects along the line-of-sight, but our point is that the
difference between the line-of-sight averaging of the ``direct''
emissions and the ``thermal'' emissions is highly
suppressed---depending only on $n_{\VM}^{-4/17}$.  This allows us to
directly compare all emissions in order to estimate the order of
magnitude for the energy budget discussed in
Section~\ref{sec:energy-budget}.  Once the other calculations and
observations are better constrained, one might be able to search for
this type of scaling to test our proposal more rigorously.  \clearpage
\end{document}